# On Degree-Based Decentralized Search in Complex Networks

Shi Xiao    Gaoxi Xiao

Division of Communication Engineering
School of Electrical and Electronic Engineering
Nanyang technological University, Singapore 639798
Email: shixiao@pmail.ntu.edu.sg, egxxiao@ntu.edu.sg

*Abstract*— Decentralized search aims to find the target node in a large network by using only local information. The applications of it include peer-to-peer file sharing, web search and anything else that requires locating a specific target in a complex system. In this paper, we examine the degree-based decentralized search method. Specifically, we evaluate the efficiency of the method in different cases with different amounts of available local information. In addition, we propose a simple refinement algorithm for significantly shortening the length of the route that has been found. Some insights useful for the future developments of efficient decentralized search schemes have been achieved.

## I.  INTRODUCTION

Recently there has been increasing research interest in studying large-scaled real-world systems by formulating them into various network models. It is found that different systems, varying from Internet, world-wide web (WWW), airline transportation systems, scientific co-authorship, food web, to protein-protein reactions, even terrorism organizations, when formulated into network models, share some stunning common features. The above observations have spawn a new area called *complex networks* [1, 2].

One of the most important complex-network models is the *scale-free network* [1, 2]. In such networks, the nodal-degree distribution (i.e., the fraction of nodes with *k* connections) behaves as:

$$P(k) \propto k^{-\alpha}, \ m \leq k \leq M$$

where $\alpha$ is the exponent, *m* is the lower cutoff, and *M* is the upper cutoff. It is claimed (though not without arguments) that all the systems we mentioned above can be formulated as the scale-free networks [1, 2].

An important operation in complex networks is *decentralized search*, i.e., the operation of locating a certain target without a central controller to provide global information. The applications include peer-to-peer file sharing as that in Gnutella and Napster [3, 4], web crawlers search [5, 6], and anything else that requires locating a specific target in a complex system. The methods without considering the special features of the corresponding network models include breadth-first searching methods based on limited flooding [7, 8], random walks [9], and various *informed searching* methods with available information ranging from simple forwarding hints to exact object locations [10, 11]. By making use of the special features of the system topologies, most notably the "six-degree separation" in small-world networks [12], the well-known methods include the degree-based searching [13, 14] and the similarity-based searching methods [15-17].

The degree-based search methods typically make the assumptions that (i) each node knows its own neighborhood network topology; and (ii) each node can locate the target if and only if the target is within a certain range of its neighborhood. If the target is not found, the request will be forwarded to a chosen set of adjacent nodes for further search. In the degree-based method, generally speaking the request is forwarded to high-degree nodes. In fact, the most typical method simply forwards the request to the highest-degree adjacent node [14].

The similarity-based method, on the other hand, explores the similarity features existing in many real-world systems. For example, people like to watch men's basketball games may know more about women's basketball as well. Therefore it may be a good idea to ask for information

This work was partly supported by Microsoft Research Asia (MSRA).

about some female basketball players to a community of men's basketball fans. From an algorithm-development point of view, however, it can be highly tricky to develop proper measurements of similarity, which sometimes have to be application dependent [16, 17]. Recently, there are research efforts aiming to combine the degree-based and similarity-based searching methods [13]. The results turn out to be quite encouraging.

We in this report go back to study the very fundamental degree-based methods, with two major objectives: (1) to better understand how the performance of such methods can be affected by the amount of the local information being available to each node; (2) to develop a simple distributed refinement algorithm that can shorten the route between the source-destination nodes. The reason we try to shorten each route is that a shorter route generally speaking is more effective in supporting information exchanges between source-destination nodes, and usually leads to better robustness as well. Our main observations can be summarized as follows:

- Comparing the cases where each node can identify the target node within one, two or three hops away from itself respectively (We call such cases as having one-, two-, and three-hop information, respectively.), we find that the efficiency of the degree-based searching method can be drastically improved when more information is available.

- Having complete two-hop information plus partial three-hop information helps to improve the performance of the degree-based search compared to the case with only complete two-hop information. However, it does *not* easily improve the searching performance to be comparable to that of the case with complete three-hop information.

- The proposed simple refinement algorithm can shorten the average hop length of each route to be close to the theoretical *optimal* solution, i.e., the average length of the shortest paths between all the source-destination node pairs.

The rest parts of this paper are organized as follows. In Section II, we define in details the degree-based searching methods we would investigate. These methods are evaluated by extensive simulations in Section III. The refinement algorithm for shortening the route is reported and evaluated in Section IV. Finally, Section V concludes the paper.

## II. Degree-based decentralized searching methods

The first degree-based searching method we will study is the simple method proposed in [14]. Specifically, the operation procedure is as follows:

Upon the arrival of a searching request, each node will check the one-hop, two-hop or three-hop information it has. If the target is identified among its neighborhood, the search stops; otherwise, there are two different cases:

1) If some nodes adjacent to the current node have never been visited by the searching request, the searching request will be forwarded to the largest-degree node among them. When there is a tie, break it randomly.

2) If no such adjacent node exists (In other words, all the adjacent nodes have been visited by the searching request.), the searching request will be deflected back to the node where it came from.

The searching operation will be terminated if the request has been forwarded/deflected many times yet still cannot find the target. In our simulations, we terminate the search when the number of forward/deflection operations equals to the number of nodes in the network, which seldom happens in our experiences.

A slightly different method is reported in [14]: in case 2), if all the adjacent nodes have been visited by the searching request before, we find among them those nodes whose single-hop neighborhood has not been *exhaustively* visited by the searching request. Then we forward the request to the highest-degree node among them. Our experiences show that the two methods perform nearly the same. The latter one only occasionally helps to shorten the searching procedures. Therefore in this paper, we discuss the former method only.

As aforementioned, we will evaluate the performance of the method with one-hop, two-hop and three-hop information, respectively (while in [14], it has always been assumed that the two-hop information is available). In many real-world systems, however, it is not easy to acquire complete two-hop information, let alone three-hop information. For example, in the Internet, a big hub may have hundreds of thousands of neighborhood nodes within three-hop distance from itself. To test the possibility of loosening the request of having complete three-hop information without significantly sacrificing searching performance, we consider the following *partial three-hop information-based* (Partial-3) searching method:

If the target cannot be found by a node within its two-hop neighborhood, this node will "consult" a few adjacent nodes whether the target is within their two-hop neighborhood. If a positive reply comes back, the search stops; otherwise, the search request is forwarded (or deflected) in the same way as that in the original method we discussed earlier. In our simulations, we let each node to consult its largest-degree adjacent nodes that have never been consulted before.

### III. EVALUATIONS OF THE EFFECTS OF DIFFEERNT INFORMATION AVAILABILITIES

We first evaluate the performance of the simple method proposed in [14] with one-, two-, and three-hop information, respectively. Numerical simulations are conducted on the Barabási-Albert (BA) model [1] containing 10,000 nodes and a real-world Internet model containing 6,470 nodes [18]. In each model, we simulate decentralized search between 500 pairs of randomly chosen source-destination nodes. The average results of 10 rounds of such simulations are presented in Fig. 1.

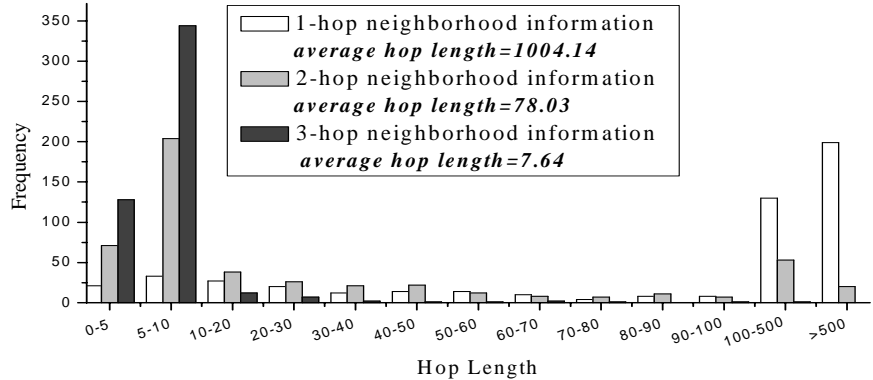

(a)   BA model

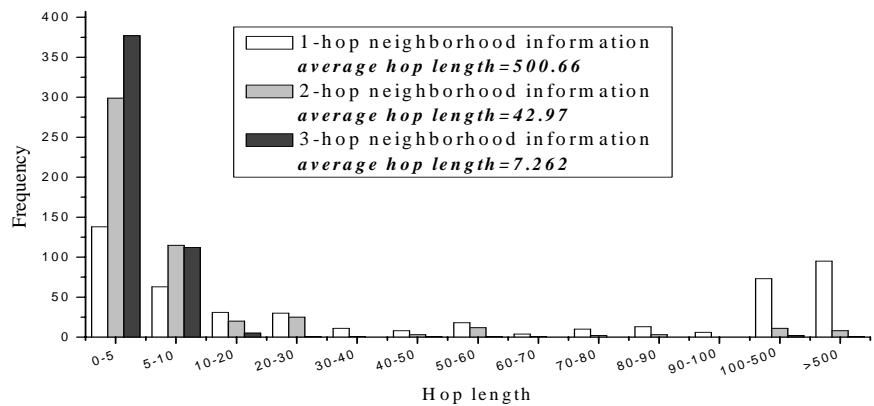

(b)   6470-node Internet model

Fig. 1. Performance of the classic degree-based searching method with different amounts of local information.

The simulation results evidently show that having more local information helps to drastically improve the efficiency of the searching operations. For example, in the BA model, with single-hop information, the average number of hops each search operation goes through is as large as 1004.14. This number can be reduced to 78.03 when two-hop information is available; and further reduced to 7.64 when three-hop information is available. The numbers of those search operations going through fewer than 10 hops account for 10.8%, 55%, and 94.4% among all the 500 operations in the three different cases, respectively. Similar conclusions hold in the Internet model.

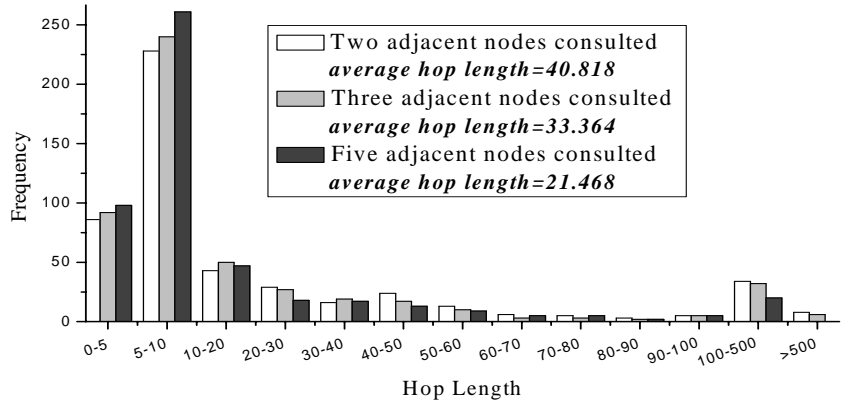

(a) BA model

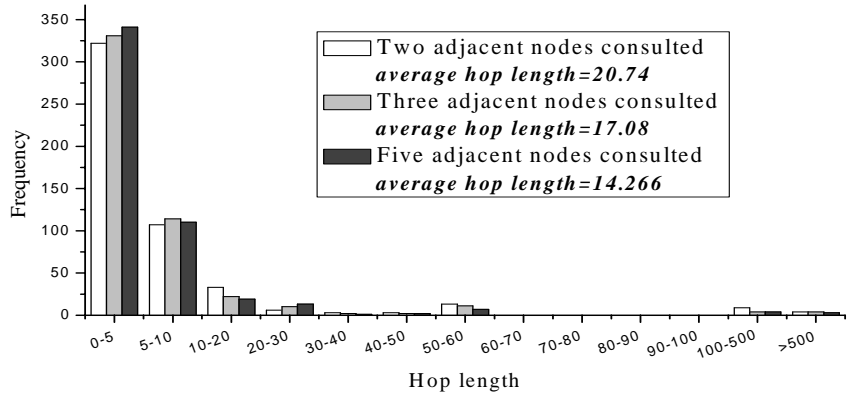

(b) 6470-node Internet model

Fig. 2. Performance evaluations of the Partial-3 method.

The results in Fig. 1 show that, having complete two-hop information is not sufficient to achieve comparable performance to that of the centralized search (where the shortest path between each node pair can be calculated). Considering the fact that it may not be feasible to acquire three-hop information in many real-world systems, we test the Partial-3 method where each node can consult at most 2, 3 or 5 of its adjacent nodes respectively. The simulation results are presented in Fig. 2. We see that compared to the case with complete two-hop information, the Partial-3 method performs much better. However, it does not improve the performance to be close to that of the case with complete three-hop information. In the BA model, even when each node can consult 5 adjacent nodes, the average hop length of each search operation is still 21.468 (With complete three-hop information, the average path length would be a much shorter 7.64.). The high sensitivity of searching performance to the amount of available local information may imply necessity of combining the degree-based and the similarity-based searching methods: searching based on network topology information (such as nodal-degree information) alone may simply never be good enough in certain circumstances.

## IV. REFINEMENT METHOD AND ITS PERFORMANCE

For many applications, finding the target is not the end of the story. Once the target is found, a path needs to be set up between the source and the target. From the discussions in the last section, we see that, unless we have extensive three-hop information, the searching request may have to go through a long path before it reaches the destination. If a short path between the source-destination nodes cannot be found at the first place, for many applications, it is helpful to shorten the path afterwards. For example, for peer-to-peer multimedia file sharing, some delay in connection setup generally speaking can be tolerated; yet it is important to make sure that the set-up connection is short, robust and with sufficient bandwidth. Motivated by such applications, we propose the following simple distributed refinement method with the objective of shortening a route that has been found.

Assume the searching request finally reaches the target node carrying a list of all the nodes it has passed through. Denote the node list as $\{s, n_1, n_2, ..., n_k, t\}$, where $s$ is the ID of the source, $t$ the ID of the target, and $n_i$, $i = 1, 2, ..., k$ the intermediate nodes. Find among the node list the *first* one that is adjacent to the destination. Denote this node as node *m*. Then find among the node list the first one adjacent to node *m*. Repeat the procedure under the node *s* is finally reached.

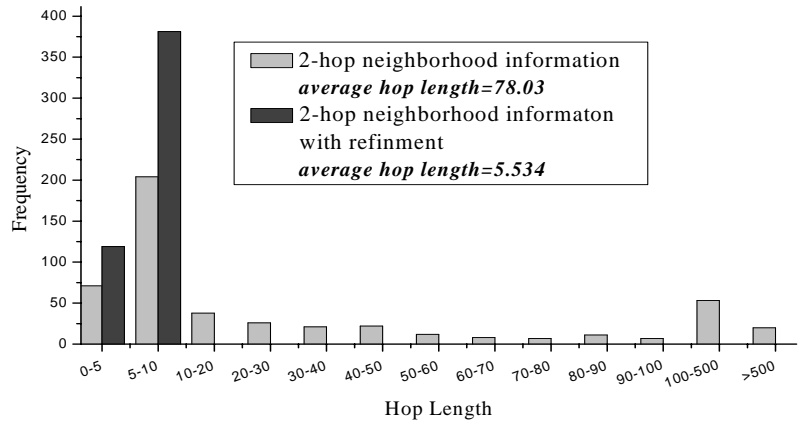

(a) BA model

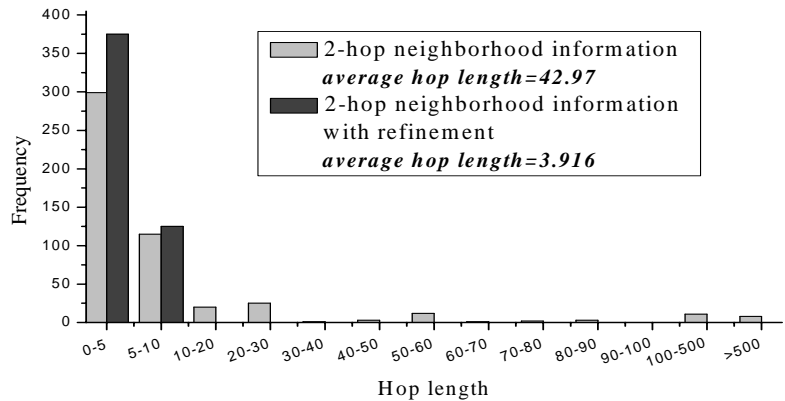

(b) 6470-node Internet model

Fig. 3. Performance evaluations of the refinement method.

Apparently, the refinement method adopts a simple greedy approach to take "shortcuts" of the route that the searching request has gone through. The performance of the method is presented in Fig. 3, where we assume that the complete two-hop information is available to every node. Simulation results show that, simple as it is, the refinement method drastically shortens the average hop length of each path from 78.03 to 5.53 in the BA model, close to the average length of the shortest paths between all the node pairs (which is 4.88). More significantly, not a single path after refinement has more than 10 hops. Therefore, in applications where refinements are allowed, the simple algorithm can effectively achieve close-to-optimal solutions.

## V. Conclusion

We studied the degree-based decentralized searching methods in complex networks. Extensive simulations results show that the performance of such methods is rather sensitive to the amount of available local information. Such high sensitivity cannot be easily eliminated by limited information exchanges between neighborhood nodes, even if these nodes are of high degrees and are in scale-free networks with relatively large numbers of hub nodes. However, by using a simple refinement algorithm, we can shorten most of the routes to be close to the shortest lengths they could ever achieve. These insights would be helpful for the future developments of more efficient decentralized searching methods.

## References

[1] S. Bornholdt, H. G. Schuster (Eds.), *Handbook of Graphs and Networks: From the Genome to the Internet,* Wiley-VCH, 2003.
[2] E. Ben-Naim, H. Frauenfelder, and Z. Toroczkai (Eds.), *Complex Networks*, SpringerVerlag Berline Heidelberg, 2004.
[3] Gnutella website: http://www.gnutella.com
[4] Napster website: http://www.napster.com
[5] S. Chakrabarti, M. van den Berg, and B. Dom, "Focused Crawling: A New Approach to Topic-specific Web Resource Discovery," In *Proceedings of the $8^{th}$ International World Wide Web Conference*, 1999.
[6] M. Diligenti, F. Coetzee, S. Lawrence, C. L. Giles, and M. Gori, "Focused Crawling Using Context Graphs," In $26^{th}$ *International Conference on Very Large Database (VLDB)*, 2000.
[7] V. Kalogeraki, D. Gunopulos, and D. Zeinalipour-Yazti, "A Local Search Mechanism for Peer-to-Peer networks," In *Proceedings of the 11th international Conference on Infomation and Knowledge Management (CIKM)*, 2002.
[8] B. Yang and H. Garcia-Molina, "Improving Search in Peer-to-Peer Networks," In *IEEE International Conference on Distributed Computing Systems (ICDCS)*, 2002.
[9] C. Lv, P. Cao, E. Cohen, K. Li, and S. Shenker, "Search and Replication in Unstructured Peer-to-Peer Networks," In *Proceedings of the 16th annual ACM International Conference on supercomputing (ICS)*, 2002.
[10] D. Tsoumakos and N. Roussopoulos, "Adaptive Probabilistic Search (APS) for Peer-to-Peer Networks," Technical Report CS-TR-4451, University of Maryland, 2003.
[11] A. Crespo and H. Garcia-Molina, "Routing Indices for Peer-to-Peer Systems," In *IEEE International Conference on Distributed Computing Systems (ICDCS)*, 2002.
[12] J. Travers and S. Milgram, "An Experimental Study of the Small World Problem," *Sociometry*, 32, 1969.
[13] O. Simsek and D. Jensen, "Decentralized Search in Networks Using Homophily and Degree Disparity," In *Proc. 19th International Joint Conference on Artificial Intelligence*, 2005.
[14] L. A. Adamic, R. M. Lucose, A. R. Puniyani, and B. A. Huberman, Search in power-law networks, *Physical Review E*, 64, 2001.
[15] J. Kleinberg, "Navigation in a small world," *Nature*, 406:845, 2000.
[16] J. Kleinberg, "The Small-World Phenomena: An Algorithmic Perspective," In *Proceedings of the $32^{nd}$ ACM Symposium on Theory of Computing*, 2000.
[17] J. Kleinberg, "Small-world Phenomena and the Dynamics of Information," In *Advanced in Neural Information Processing Systems (NIPS)*, vol. 14, 2001.
[18] http://moat.nlanr.net/Routing/rawdata